# Living Cell Cytosol Stability to Segregation and Freezing-Out: Thermodynamic aspect

Viktor I. Laptev
*Russian New University, Moscow, Russian Federation*

The cytosol state in living cell is treated as homogeneous phase equilibrium with a special feature: the pressure of one phase is positive and the pressure of the other is negative. From this point of view the cytosol is neither solution nor gel (or sol as a whole) regardless its components (water and dissolved substances). This is its unique capability for selecting, sorting and transporting reagents to the proper place of the living cell without a so-called "pipeline". To base this statement the theoretical investigation of the conditions of equilibrium and stability of the medium with alternative-sign pressure is carried out under using the thermodynamic laws and the Gibbs" equilibrium criterium.

## I. INTRODUCTION

Cytosol in a living cell (intracellular fluid or cytoplasmic matrix, hyaloplasm matrix, aqueous cytoplasm) is a combination of the water dissolved substances. It places in the cell between the plasma membrane, the nucleus and a vacuole; it is a medium keeping granular-like and whisker-like structures. Although water forms the large majority of the cytosol, its structure and properties within cells is not well understood. The cytosol is always moving, it has about a half of the volume of the cell and one third of the cell mass. It supports the chemical transport, takes part in glycolysis, synthesis and destroying amino acids and in the protein (albumen) synthesis. The cytosol is a homogeneous solution and, at the same time, it is a system containing the interacted parts [1-4]. As is known, the Gibbs' phase rule explains freezing-out and segregation stability of sea water, but the unique cytosol stability to segregation and freezing-out has no thermodynamic explanation until recently.

## II. KNOW-HOW, BASIC THERMODYNAMIC DIFINITIONS

The paper reports the conditions of equilibrium and segregation stability of a medium consisting of a positive pressure phase and a negative pressure one. It is shown that this unusual state of the medium with the negative pressure and positive temperature is not metastable and coexists without segregation in equilibrium with an usual medium state where the pressure and temperature are positive.

The considered medium is a matter as a homogeneous or a heterogeneous thermodynamic system. We suppose that the homogeneous medium can exist as a combination of two phases without segregation. This state is called homogeneous phase equilibrium. In this paper the properties of multiphase homogeneous medium are described.

### A. Inertial Motion in $U,S,V$-Space

A full description of the thermodynamic state of a medium without chemical interactions is given by a relationship between the internal energy $U$, entropy $S$ and volume $V$ [5]. The mathematical procedure for a negative pressure supposes using absolute values of the internal energy $U$ and entropy $S$.

The surface $\varphi(U,S,V) = 0$ corresponds to the all equilibrium states of the medium. The isolated medium spontaneously reaches the thermodynamic equilibrium state at any point of this surface. The internal forces will bring the medium to equilibrium if the entropy will be maximized and the internal energy will be minimized [5 - 8].

This work supposes that the impulse of the internal forces is arisen in the medium during a time period of its motion to the equilibrium; this pulse in conserved under compensation of external forces or their absence. In this case instead of the rest state the inertial motion of the medium without work production is possible.

### B. Zero and Negative Pressure

The inertial motion of the medium is described by the trajectory in the surface $\varphi(U, S, V) = 0$ which can intercept the regions of positive, negative or zero pressure in a general case.

The effect of the negative pressure is known as a cavitation in fluids and a homogeneous tensile deformation of a solid. The mechanics knows a hypothetical Chaplygin gas with a negative pressure. Cosmology operates with a category of negative pressure for a homogeneous isotropic medium. A cosmologic medium state with a zero pressure is called a "dust"-like one. However, we don't know an example of the medium or phase equilibrium with a sign-alternate pressure.

## III. CYTOSOL AS A THERMODYNAMIC MEDIUM

The cytosol is modeled as a homogeneous medium with a constant mass and without chemical interactions.

### A. Cytosol in Equilibrium State

The cytosol in a living cell may be considered as a homogeneous medium in equilibrium. Actually, we can concern the relationship between the following parameters: the time of pressure stabilization, temperature and water density, sound velocity ($v \sim 1400$ m/s), diffusion coefficient ($D \sim 0.2$ m$^2$/s), thermal diffusivity ($K \sim 0.561$ W/(m K)) and the linear dimension of the cell ($L \sim 20$ μm). Then the relaxation time of the cell water relating to the pressure change $L/v$ is $3 \times 10^{-9}$s, the density relaxation time $L^2/D$ is $2 \times 10^{-11}$s and the temperature relaxation time is $L^2/K \sim 8 \times 10^{-12}$s. The relaxation times in a water solution or cytosol are somewhat larger [9]. But in this case the time of metabolic processes is considerably exceeding the cytosol relaxation times to the equilibrium, i.e. these times are quasi-static.

### B. Cytosol under Inertial Motion

The condition of the inertial motion of cytosol can be realized in a living cell. There are no principal contradictions to the statement that the cytosol in a living cell moves on expense of internal resources. Really, the cytosol is surrounded by the plasma membrane and doesn't experience the deviations of the atmospheric pressure. Osmosis produces in the cell pressure which is compensated by a pressure of the cell wall. There neither hydrostatic pressure drop nor temperature gradient in the cytosol and the cell.

### C. Cytosol as a System with Finite Energy

The fundamental concepts of energy and entropy for ecologic and biologic systems concern the living beings as open non-equilibrium thermodynamic systems permanently exchanging energy and matter with the environment [10-12]. We make an emphasis on the fact that the energy resources and entropy of the living organism are finite. So the cell as the basic structural and functional unit of all known living organisms is concerned as a thermodynamic system with limited internal energy.

## IV. THEORY

The first law of thermodynamics for systems with finite energy is unchanged as a law about entropy existing and increasing [8]. Pressure $p$ and temperature $T$ of the cytosol are determined by a plane tangent to the surface $\varphi(U,S,V)=0$ of our model medium. It is an algebraic surface $\varphi_1(U,S,V)=0$ of the first order in orthogonal coordinates, its slope to the surface $\varphi=0$ determines pressure $p$ and temperature $T$ by the equations $p = -(\partial U/\partial V)_S$ and $T=(\partial U/\partial S)_V$. The coordinates of the tangent points are given by the equation

$$U = TS - pV \qquad (1)$$

after superposition of tangent surfaces at the origin of coordinates. When energy $U$ is restricted, the sign of $p$ and $T$ alternates. The system of nuclear spins of a crystal with a total energy and entropy restricted by the spin orientation can serve as an example [8, 13]. Media with finite internal energy are known as unusual thermodynamic systems.

In algebra the general condition of thermodynamic equilibrium for usual and unusual phases is as follows: the virtual displacements of total entropy $\delta S^* = 0$ under constant $U^*$ and fixed $V$ [6]. For example, the isolated system from two phases 1 and 2 with ratios $S^*=S_1+S_2$ and $U^*=U_1+U_2$, $T\delta S=\delta U+p\delta V$ and $\delta U_1+\delta U_2=0$ is described by the equation

$$(1/T_1 - 1/T_2)\delta U_1 + (p_1/T_1)\delta V_1 + (p_2/T_2)\delta V_2 = 0 \qquad (2)$$

with phase indices only.

### A. Cytosol as an Usual Medium

It is known [6-8], that if $\delta V_1 = -\delta V_2$ the solution of (2) is

$$T_1 = T_2, \qquad p_1 = p_2,$$

In geometry these usual states 1 and 2 are identified by two tangent points of the plane to the surface $\varphi(U, S, V) = 0$ [5].

### B. Cytosol as an Unusual Medium

According to (1), the negative pressure is possible only under the negative pressure. Nevertheless, the reverse speculation is not possible: one cannot state that the negative pressure exists under negative temperatures only. So, in the equilibrium thermodynamic system the alternating-sign pressure is possible under positive pressure. For example, the isolated system with $\delta V_1 = \delta V_2$ has a solution of (2):

$$T_1 = T_2, \qquad p_1 = -p_2. \qquad (3)$$

Geometrically it is presented by two tangent planes to the surface $\varphi=0$ with a total slope to the surface $U$=const is $180^0$.

### C. Medium Segregation Stability

We don't know any analysis of the stability of the alternating-sign pressure medium. Let us examine their stability related to the initial medium using the Gibbs' stability criterion [7,8]. The initial medium with constant mass is homogeneous and isotropic. Indexing its entropy $S_0=S_1+S_2$ we rewrite (1) as

$$U_0 - T_0 S_0 + p_0 V_0 = 0. \qquad (4)$$

We have to find out which of signs =, > or < appears between the internal energy $U_0$ of the initial medium and the sum $U_1+U_2$

## D. Segregation Stability of the Usual Medium

Let us consider the decay of the initial medium with formation of phases 1 and 2. If $V_1+V_2=V_0$ and $p_1=p_2$, expressions (1), (4) may be rewritten as

$$U_0 - (p_{1,2} - p_0) V_0 = U_1 + U_2. \quad (5)$$

Then we examine the phase equilibrium stability relating the initial medium under variation of the pressure sign in (5) for three cases.

*Scenario 1i.* If $p_0 = p_1 = p_2$, instead of (5) we have

$$U_0 = U_1 + U_2.$$

Sign = here means that the initial medium and the phase can be in equilibrium under segregation. To visualize our description we denote the internal energy as $\bar{U}$ when the pressure is negative. Then the sign = in the expression below

$$\bar{U}_0 = \bar{U}_1 + \bar{U}_2$$

means that the initial medium and phases can be in equilibrium under negative pressure.

*Scenario 2i.* When $p_0 = 0$ and $p_1 = p_2 > 0$, instead of (5) we have

$$U_0 > U_1 + U_2.$$

The sign > means that the initial medium is unstable relating to the equilibrium of new phases with a positive pressure under their segregation.

*Scenario 3i.* If $p_0 = 0$ and $p_1 = p_2 < 0$, instead of (5) one has

$$U_0 < \bar{U}_1 + \bar{U}_2.$$

The sign < means that under segregation of 1 and 2 with negative pressure their equilibrium is unstable relating to the initial medium with a zero pressure.

## E. Segregation Stability of an Unusual Medium

Now we ascribe the positive pressure to the phase 1 and the negative pressure to the phase 2. In the geometry of the $U$, $S$, $V$ – space it means the intercept of two planes with an unique tangency with the surface $\varphi = 0$. The interception line is an isotherm of the plane $V$=const. Then the equilibrium condition (3) is physically valid under absence of the phase segregation and $V_1 = V_2 = V_0$, where $V_0$ is a volume of the medium. In this case, according to (1), (3), (4),

$$U_0 + p_0 V_0 = U_1 + U_2. \quad (6)$$

Now we again investigate the phase equilibrium condition relating to the initial medium under variation of the pressure of initial medium $p_0$ in three cases below.

*Scenario 1j.* When $p_0 = 0$, instead of (6) we have

$$U_0 = U_1 + \bar{U}_2.$$

The sign = means that under $p_0 = 0$ the phases 1 and 2 may be in equilibrium with the initial medium without segregation.

*Scenario 2j.* If $p_0 > 0$, instead of (6) we have

$$U_0 < U_1 + \bar{U}_2.$$

The sign < means that the phases 1 and 2 are unstable relating to the initial medium with a positive pressure without segregation.

*Scenario 3j.* When $p_0 < 0$, instead of (6) we have

$$\bar{U}_0 > U_1 + \bar{U}_2.$$

The sign > means that the initial medium with a negative pressure is unstable relating to the equilibrium of new phases without segregation.

## F. Compensation of Usual and Unusual Media

*I-j* scenarios have a different energy balance for the initial medium and new phases and suppose compensation. Assume that the initial medium spontaneously decays according to the scenario 3j. Corresponding to (5), the uncompensated energy is equal to the difference

$$\Delta U = \bar{U}_0 - (U_1 + \bar{U}_2) = -p_0 V_0.$$

When the initial medium spontaneously decays according to the scenario 2i, we have $\Delta U = U_0 - (U_1 + U_2) = p_{1,2} V_0$. The zero difference $\Delta U = 0$ corresponds to the scenario 1i or 1j. The uncompensated energy $\Delta U$ is spent on the medium mechanical reconstruction at constant $p$ and $T$.

## G. Homogeneous and Heterogeneous Phase Equilibrium

The Gibbs' conditions for equilibrium and stability of thermodynamic media with an alternating-sign pressure found out in this work are allowing building two rules:
1. The equilibrium phases are stable to the segregation under inertial motion or in rest if they have pressure of opposite signs.
2. The equilibrium phases exist separately under the inertial motion or in rest if they have a pressure of the same sign.

A zero isobar from the scenarios *i-j* is a boundary curve between the homogeneous and heterogeneous phase equilibrium regions on the surface $\varphi\ (U,\ S,\ V) = 0$. Corresponding to (1), at the each point of the zero isobar the product of temperature and $U/S \equiv 1/\alpha$ equals to unity. According to the rules 1 and 2, the medium under inertial motion intercepts the zero isobar and excludes the phase segregation by itself when $T/\alpha \leq 1$.

## V. INERTIAL MOTION AND PHASE RULE

According to (1), the inertial motion of the medium takes place above the zero isobar if $U<TS$. For the zero isobar $U=TS$ and the spontaneous phase transition is possible corresponding to scenarios 1$i$ or 1$j$. The inertial motion of the medium takes place below the zero isobar corresponding to scenario 3$j$.

Regarding (1) and Gibbs' phase rule, the one-component system in the heterogeneous equilibrium may be realized no more, than from three phases. If the phases number reaches its maximum the inertial motion of the medium finishes due to absence of the thermodynamic degrees of freedom. According to scenarios 1$i$ and Gibbs' phase rule, the inertial motion of the medium with positive pressure stops in triple point.

There is no evidence that the Gibbs' phase rule is not valid for homogeneous phase equilibrium. According to (3), the phases have different pressure and Gibbs had pointed out the constant mass ratio as a peculiarity of these phases [5]. It is also kept for an inertial motion. Regarding scenarios $j$ and Gibbs' phase rule, the one-component system in the homogeneous equilibrium has one state with zero thermodynamic degrees of freedom. According to scenario 1$j$, three phases coexist under zero pressure only. So zero isobar is two phase boundary curve with other triple point for one-component system. The inertial motion finishes in this point.

## VI. CYTOSOL AS A HOMOGENEOUS PHASE EQUILIBRIUM

### A. Cytosol and Zero Isobar

Go back to (1), the inertial motion of solution, gel or sole takes place above the zero isobar if $U<TS$. For the zero isobar $U=TS$ and the spontaneous phase transition is possible corresponding to scenarios 1$i$ or 1$j$. The new homogeneous or heterogeneous phase equilibrium is now neither gel nor sol or solution despite the fact that water and dissolved substances are its components.

The phase transition heterogeneous-homogeneous equilibrium is known in the cell physiology. The protein the coagulation of the albumen in the cytosol takes place at $42^0$C. On the other hand, the cells function up to water freezing-out under cooling down to $-89^0$C in the nature and down to $-196^0$C in the laboratory. This work supposes that the cytosol is not segregated and does not freeze out water because of its inertial motion along the zero isobar and downwards.

Let's thermodynamic surface $\varphi(U, S, V) = 0$ describes a thermodynamic equilibrium of water with a positive pressure and albumen mass with a negative pressure. For a negative pressure phase $\bar{U}>TS$ is valid. Than in the inertial motion along the zero isobar and downwards the ratios $U_1<TS_1$ and $\bar{U}_2>TS_2$ are valid. The combination of signs < and > shows an unique ability of the cytosol for a space separation, sorting and transportation of reagents to the proper place of the cell without a „pipeline".

### B. Parameter α in Biological Experiment

The condition of minimum energy at the constant entropy of the cytosol is a governor of the inertial motion. The cytosol intercepts the zero isobar at the condition $T/\alpha = 1$. Then $\alpha = 1/(42+273.15) = 0.00317$ K$^{-1}$ at $42^0$C and 0.013 K$^{-1}$ at the temperature of liquid nitrogen. From this point of view the homeostasis as a regulation of the living cell chemical activity has to support the value of the parameter $1/\alpha$ of the cytosol so that the condition $\alpha T \leq 1$ is to be valid. The negative pressure appears at scattering the medium particles and a positive pressure leads to their compression. Optical study demonstrates that the dye makes a cell wall only for a living cell in the agar-agar medium and it penetrates into the cytosol of the dead cell only. One can suggest that the negative pressure of the living cell moves out the dye. Than the cell dies due to the intercept zero isobar by the cytosol and pressure switch from negative to positive according to the scenarios 1$i$ or 2$i$. Now the dye can enter membrane bag.

The cytosol can be formed from the solution, gel or sol according to the scenarios 1$i$ → 1$j$ in various ways. The experiment based on biology of the living cell requires changing the parameter α of the initial solution in the way helping to realize a transition through the zero isobar to prevent the phase segregation. In other words, biological processes between the components may be initialized in the laboratory if one can produce homogeneous phase equilibrium instead of solution, gel or sol corresponding to (3).

### C. Radiation in Biological Experiment

The zero isobar intercept leads to a change of the positive pressure with a negative one during the energy transition from the state $U<TS$ to the state $\bar{U}>TS$ at constant $T$. Without heat transfer the internal energy of the cytosol can be enlarged by means of metabolism or photosynthesis only.

The photosynthesis efficiency in the plant cell is found out in [14]. Remembering the nature photosynthesis realized by the living green leaf we should note that the thermodynamic model of the living structure proposed in this work can be a base for further fundamental research and practical studies of the possibility of designing principally novel systems based on biological objects.

### SUMMARY

To conclude we note that there are now experimental facts which allow ascribing the negative pressure to some ions, proteins, their groups or water as the cytosol components. However, it is known [15, 16] that the osmotic pressure with an opposite sign is applied to explaining transportation of substances in a plant cell. We hope that the phenomenological investigation of the cytosol from the living cell as a model from the first thermodynamics principles will find a niche among theory and experiment for explanation of mechanisms of biochemical processes in the living cell and their realization in novel biotechnologies. Analysis of cytosol phase network requires a combination of experimental and theoretical approaches including the development and analysis of simulations and modeling.